\documentclass[,final ]{aipproc}
\layoutstyle{6x9}

\usepackage{psfig}
\usepackage{amsmath}
\usepackage{amssymb}
\usepackage{graphicx}
\usepackage{natbib}

\newcommand{\lSect}[1]{{\label{sec:#1}}}
\newcommand{\lFig}[1]{{\label{fig:#1}}}

\def\gtaprx {\lower .1ex\hbox{\rlap{\raise .6ex\hbox{\hskip .3ex
    {\ifmmode{\scriptscriptstyle >}\else
        {$\scriptscriptstyle >$}\fi}}}
    \kern -.4ex{\ifmmode{\scriptscriptstyle \sim}\else
        {$\scriptscriptstyle\sim$}\fi}}}
\def\ltaprx {\lower .1ex\hbox{\rlap{\raise .6ex\hbox{\hskip .3ex
    {\ifmmode{\scriptscriptstyle <}\else
        {$\scriptscriptstyle <$}\fi}}}
    \kern -.4ex{\ifmmode{\scriptscriptstyle \sim}\else
        {$\scriptscriptstyle\sim$}\fi}}}
\newcommand{\note}[1]{\emph{\textcolor{red}{}}}

\newcommand{\Msun}{{\ensuremath{\mathrm{M}_{\odot}}}}

\newcommand{\FIGFF}[2]{{\ref{fig:#2}{#1}}}

\newcommand{\Figure}[1]{{Figure~\FIGFF{}{#1}}}

\newcommand{\Ep}[1]{{\ensuremath{10^{#1}}}}
\newcommand{\E}[1]{{\ensuremath{\powersep\Ep{#1}}}}

\newcommand{\powersep}{{\ensuremath{\times}}}
\newcommand{\cm}{{\ensuremath{\mathrm{cm}}}}

\newcommand{\CASTRO}{\texttt{CASTRO}}
\newcommand{\KEPLER}{\texttt{KEPLER}}

\begin{document}

\title{Two Dimensional Simulations of Pair-Instability Supernovae}

\classification{97.20.Wt}
\keywords      {Stellar evolution, Massive star, Pair instability supernovae}

\author{Ke-Jung Chen}{
  address={School of Physics and Astronomy, University of
  Minnesota, Minneapolis, MN 55455}
}

\author{Alexander Heger}{address={School of Physics and Astronomy, University of
  Minnesota, Minneapolis, MN 55455
  }
}

\author{Ann S. Almgren}{
  address={Computational Research Division, Lawrence Berkeley National
  Lab, Berkeley, CA 94720}
}

\begin{abstract}
  We present preliminary results from two dimensional numerical
  studies of pair instability supernova (PSN). We study nuclear burning, 
  hydrodynamic instabilities and explosion of very massive stars. Use a 
  new radiation-hydrodynamics code, \CASTRO.
\end{abstract}

\maketitle


\section{Introduction}
\label{intro}
The first stars that formed after the big bang may have had a 
characteristic mass scale about hundred times heavier than modern 
stars \citep{abel,bromm}. Stars with initial mass between $140\,\Msun$ and 
$260\,\Msun$ end their lives in a very powerful explosion, a pair-instability 
supernova (PSN) \cite{barkat,bond}. Such supernovae could play an important role 
in the synthesis of the first heavy elements \citep{heger1}. The energy
output into their surroundings can affect the structure and evolution 
of the early universe. Current theoretical models 
of PSN are mostly based on one-dimensional calculations \cite{heger1}. 
Until now, multidimensional simulations have been scarce, however. Here we present 
2D simulations that aim to study how fluid instabilities affect the 
nucleosynthesis and energetics of PSN. 

\section{Numerical Approach}
\lSect{castro} We start our simulations using one-dimensional models 
obtained from the \KEPLER{} code \cite{weaver}, an implicit spherically symmetric 
Lagrangian hydrodynamics code. We follow the 1D stellar evolution until 
$10$ secs before maximum compression. Then we map the resulting
1D profiles into 2D using a conservative mapping procedure 
\cite{ken} to serve as the initial conditions for a new Eulerian AMR code, \CASTRO{} \cite{ann1}. 
We follow the contraction, burning, and onset of explosion for about $100$ secs. During this 
period, thermonuclear burning releases almost all 
of the explosion energy.


\section{Results and Discussion}
The results presented here are from a $150\,\Msun$ star for which we simulate one hemisphere 
using cylindrical symmetry. \Figure{PSN-e} shows the oxygen mass 
fraction in the inner $2\E{10}\,\cm$ domain at about $60$ secs after maximum 
compression. We find that Rayleigh-Taylor (RT) instabilities develop at 
the outer edge of the oxygen-burning shell. The RT instabilities at $1.8\E{10}\,\cm$ are 
generated by helium shell burning. Later, we expect that these instabilities grow 
further after passing of the reverse shock. They could lead to large-scale mixing and affect 
the observable PSN light curve. Anisotropic ejection of material of different compositions 
may affect the chemical evolution of its surrounding.

\begin{figure}
\includegraphics[height= 8. cm]{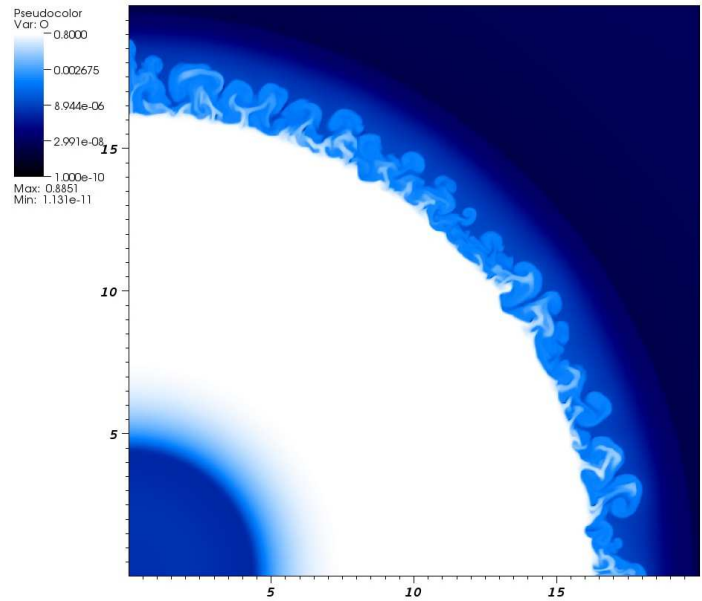}
\vspace{-\baselineskip}
\caption{Oxygen mass fraction; $r$ and $z$ coordinate are in the unit of $1\E{9}\,\cm$ \lFig{PSN-e}}
\end{figure}




\begin{theacknowledgments}
  The authors would like to thank members of the Center for
Computational Sciences and Engineering (CCSE) at LBNL for their
invalueable support with using \CASTRO{}. The simulations were performed at Minnesota
Supercomputer Institute.  This project has been supported by the
DOE SciDAC program under grant DOE-DE-SC0002300.

\end{theacknowledgments}



\bibliographystyle{aipproc}   

\bibliography{paper1_bibliography}

\end{document}